Applic.Analysis, 59, (1995),377-383.
 \input amstex
\documentstyle{amsppt}
%\NoPageNumbers
 
\hsize32pc
\vsize50pc
 
\hcorrection{.5 in}
\vcorrection{1.0 in}
 
\TagsOnRight
 
\topmatter
\title Uniqueness theorems for inverse obstacle  scattering problems in
 Lipschitz domains 
\footnotemark"*"
\endtitle \footnotetext "*"{This work was done under the auspices of
DOE. \hfill}
\rightheadtext{Uniqueness theorems}
\author A.G. Ramm \endauthor
\affil  Applied Mathematics Department, Complutense University, Madrid
 28040; Department of
 Mathematics, Kansas State University,
        Manhattan, KS  66506-2602 \endaffil
\thanks The author thanks  LANL for support and A.Ruiz and D.Yakubovich
 for useful discussions.
 \endthanks
 
\abstract {For the Neumann and Robin boundary conditions the uniqueness
 theorems for inverse
 obstacle scattering
problems are proved in Lipschitz domains. The role of non-smoothness of
 the boundary is
 analyzed.  }\endabstract
 
\endtopmatter
 
\vglue .2in
 
\document
 
\subhead 1. Introduction \endsubhead

Let $D \subset \Bbb R^n,\,n\geq 2,$ be a bounded domain with a sufficiently
 smooth boundary
 $\Gamma$, not necessarily connected, but consisting of a finitely many
 connected components. Let
 $ D^\prime:= \Bbb R^n\setminus D$ be the exterior domain,  $k>0$  a fixed
 wavelength,
 $\alpha\in S^{n-1}$  a given unit vector, $S^{n-1}$ the unit sphere.  It
 is well known that the 
 obstacle scattering problem:
$$
\triangledown^2 u+k^2u=0\text{ in } D^\prime,\tag 1
$$
$$
u_N=0\text { on }\Gamma,\tag 2
$$
$$
u=u_0+v,\quad u_0:=\exp (ik\alpha\cdot x),\tag 3
$$
where  $v$ satisfies the radiation condition
$$
\underset r\rightarrow\infty\to{\ lim}\int_{|x|=r}|v_r-ikv|^2ds=0, \tag 4
$$
and $N$ is the exterior unit normal to $\Gamma$
has been studied intensively and there are many ways known for
 proving the existence
 and uniqueness of its solution which is called the scattering solution [3].
 The function $v$ has
 the following asymptotics
$$v=A(\alpha^\prime, \alpha,k) \gamma(r) +o(\frac 1r) \text{ as }
 r\rightarrow \infty,\,
 \alpha^\prime:=x/r.     \tag 5
$$
The coefficient $A(\alpha^\prime, \alpha,k)$ is called the scattering
 amplitude.

We also consider the Robin boundary condition in place of (2):
$$
u_N+ \sigma(s)u=0\text { on }\Gamma,\tag 6
$$
where $\sigma$ is a continuous real-valued function on $\Gamma$.

In what follows, we denote by a subindex zero the quantity which is fixed.
 The inverse obstacle
 scattering problems (IOSP1-5) can be stated as follows:

1) Given $A(\alpha^\prime, \alpha_0,k)\,\,\forall \alpha^\prime\in S^{n-1},
\forall k\in [a,b], \, 0\leq a<b ,$ find $\Gamma$, or, if Robin's
condition is assumed, find $\Gamma$ and $\sigma$;

2)  Given $A(\alpha^\prime, \alpha,k_0)\,\,\forall \alpha^\prime,\alpha 
\in S^{n-1},$ find
 $\Gamma$, or, if Robin's
condition is assumed, find $\Gamma$ and $\sigma$;

3) Given $A(\alpha^\prime, \alpha_0,k_0)\,\,
\forall \alpha^\prime\in S^{n-1},$ find $\Gamma$, or,
 if Robin's
condition is assumed, find $\Gamma$ and $\sigma$;

4) Given $A(-\alpha, \alpha,k_0)\,\,\forall \alpha\in S^{n-1},$ find
 $\Gamma$, or, if Robin's
condition is assumed, find $\Gamma$ and $\sigma$; (backscattering data)

5) Given $A(-\alpha, \alpha,k)\,\,\forall \alpha\in S^{n-1},
\forall k\in [a,b], \, 0\leq a<b ,$ find $\Gamma$, or, if Robin's
condition is assumed, find $\Gamma$ and $\sigma$;

Of course, if IOSP4 is solved then IOSP5 is solved. 

In all these problems one can assume for uniqueness studies that the
data are given on open subsets of $S^{n-1}$, however small, since
such data allow one to uniquely recover the data on all of  $S^{n-1}$ [3].

In this paper we discuss only IOSP1-2. Uniqueness of the solution to other
 three problems has been
 (and still is) an open problem for several decades, although for IOSP5
 uniqueness for convex
 obstacles follows from the results in [3].
The reconstruction of $\Gamma$ from the scattering data is not discussed here,
see [3],[4] and references therein.
 
The history and various proofs of the uniqueness theorems for IOSP1,2 are
 given
in [3],[4] and references therein, and a new method of proof and its
 applications are given in 
[5]-[8].

Uniqueness of the solution for IOSP1 was proved by M.Schiffer (1962) for the
Dirichlet boundary condition, while for IOSP2 it was proved by A.G.Ramm (1985)
for the Dirichlet, Neumann and Robin boundary conditions  (see  [3] for
 these proofs). In [5]-[7] a
 new method of proof was given.

  In this paper we discuss the technical question: the role of smoothness
 of the boundary in the
 various proofs of the uniqueness theorems for IOSP1-2. We justify the
 applicability of Green's 
formula in the 
Schiffer's and other proofs and point out that the question of whether
 the Neumann Laplacian has a
 discrete spectrum in a certain domain with non-smooth boundary can be
 avoided completely. This
 question arises in the Schiffer's type
of proofs.

Furthermore, we generalize the uniqueness results for Lipschitz domains,
 i.e., for domains with 
Lipschitz boundaries.

In section 2 the Schiffer's type proof and the proof from [6],[7] are
 presented,
and the role of the non-smoothness of some of the domains, used in these
 proofs,
is analyzed. An important role is played by the sets of finite perimeter and 
Green's formula for such sets. The related theory is discussed in [1],[2]
 and [9].
 
 We first assume in this paper that the boundary $\Gamma$ is sufficiently
 smooth and then show that
 our argument is valid in Lipschitz domains. So, this paper  deals with the
 technical problems.

Recall that a Lipschitz domain is a bounded domain each point of whose
 boundary has a neighborhood
 in which the equation of the boundary in the local coordinates is given by
 a function satisfying
 a Lipschitz condition. Lipschitz
domains are denoted as $C^{0,1}$ domains. In [10] the potential theory
results are given for Lipschitz domains.

The definition of the solution to problem (1)-(3) in non-smooth domains 
is as follows:

A function $u\in H^2_{loc}\cap H^1(D^\prime _R)$ solves (1)-(3), iff it
 satisfies conditions (3)
 and (4), and the following identity:
$$
\int_{D^\prime}(k^2u\phi -\triangledown u\triangledown \phi)dx=0\quad
\forall \phi \in  H^2_{loc}\cap H^1_c(D^\prime ).   \tag 7
$$
Here $H^l$ is the Sobolev space, $H^2_{loc}$ is the space of functions which 
are in $H^2(\tilde D^\prime)$ for any compact strictly inner subdomain 
$\tilde D^\prime$ of
 $D^\prime,\, H^1_c(D^\prime)$ is the space of functions which vanish near
 infinity (but not
 necessarily near $\Gamma$), and $H^1(D^\prime _R)$ is the 
space
of functions which for any sufficiently large $R$ belong to 
$H^1(D^\prime \cap B_R)$, where $B_R$ is the ball of radius $R$, 
centered at the origin.
This definition does not require any smoothness of the boundary.

The solution to (1), (6), (3) is a function in  
$H^2_{loc}\cap H^1(D^\prime _R)$
which satisfies conditions (3) and (4), and the identity

$$
\int_{D^\prime}(k^2u\phi -\triangledown u\triangledown \phi)dx +
\int_{\Gamma}\sigma u\phi ds=0\quad
\forall \phi \in  H^2_{loc}\cap H^1_c(D^\prime ).   \tag 8
$$
Here the Lipschitz boundary $\Gamma$ is admissible because the 
imbedding theorem
holds for such a boundary.

 In this paper we use the following notations: 
$D_{12}:=D_1\cup D_2,\, D^{12}:=D_1\cap D_2,\, \Gamma_{12}$ is the
 boundary of 
 $D_{12}$, $\Gamma^{12}$ is the boundary of $D^{12}$, 
$\Gamma_1^\prime$ is the part of $\Gamma_1$ which lies outside of $D_2$,
 and  $\Gamma_2^\prime$
is defined likewise, $\tilde D_1$ is a connected component of 
$D_1\setminus D^{12}$, $D_3:=D_{12}\setminus D^{12}$.

\vskip .2in
 
\subhead 2. Uniqueness results for IOSP with the Neumann and Robin boundary
 conditions
  \endsubhead
 
\subsubhead 2.1.Uniqueness for IOSP1\endsubsubhead 

Consider IOSP1 first. Let us outline a variant of the Schiffer's type of
 proof, which allows us
 to deal with non-smooth boundaries of the domains arising 
in the proof. Assume that there are two different obstacles, $D_j, j=1,2,$
which generate the same scattering data for IOSP1. Let $w:=u_1-u_2$, where
$u_j$ are the corresponding scattering solutions. The function $w$ solves
 equation (1) in
  $D_{12}^\prime$ and $w=o(1/r)$ because the scattering data 
are the same for $D_1$ and $D_2$. Thus, lemma [3,p.25] implies $w=0$ in
 $D_{12}^\prime$. Let $U:=u_1=u_2$ in  $D_{12}^\prime$. Then $U$ can be
 continued analytically, 
as a solution to (1) , to the domains  $D_3$ and $(D^{12})^\prime$,
because either $u_1$ or $u_2$ are defined in these domains and solve (1)
 there. 
We assume that $(D^{12})^\prime$ is not empty. If it is, the argument is even
 simpler: $V:=U-u_0$
 solves equation (1) in $\Bbb R^n$ and satisfies the radiation condition;
 thus, $V=0$ and $U=u_0$
 in $\Bbb R^n$. Since $u_0$ does
not satisfy the boundary condition (2), we have got a contradiciton. 
This contradiction proves that the assumption  ($(D^{12})^\prime$ is empty) is
wrong.

The domain $D_3$ is bounded since both $D_j$ are. The function $U$ solves 
equation (1) and satisfies the homogeneous boundary condition (2) on
its boundary $\Gamma_3$, except for, possibly, the set of $(n-1)$-dimensional
Hausdorff measure, namely, except for the set of points which belong to the
 intersection of
 $\Gamma_1$ and $\Gamma_2$. Since the scattering solutions 
in domains with smooth boundaries are uniformly bounded functions whose
 first derivatives are
 smooth (Lipschitz are sufficient for our argument), the function $U$ has
 the same properties. 
Therefore, for any $k\in [a,b]$, $U$ is in $L^2(D_3)$, and, as we prove 
below, the functions $U$, corresponding 
to different $k$, are orthogonal in $L^2(D_3)$. Since this Hilbert space 
is separable, we arrive
 at a contradiction: existence of a continuum of orthogonal non-trivial
 elements
in the separable Hilbert space $L^2(D_3)$. This contradiction proves that
 $D_1=D_2$, 
and the uniqueness theorem is proved for IOSP1. The original Schiffer's 
argument presented in the literature, uses discreteness of the spectrum of
 the Laplacian,
corresponding to a boundary condition, in a bounded domain. The discreteness
 of the spectrum holds for any bounded domain for the Dirichlet Laplacian,
 but not necessarily for the Neumann one. This is why we want to avoid the
 reference to the
 discreteness of the spectrum of the Neumann Laplacian or 
the Robin Laplacian.
To complete the proof, it is sufficient to prove the claim about the
 orthogonality of $U$ with
 different $k$. The proof of this claim goes along 
the usual line. The new point is the discussion of the applicability of
 Green's formula, used
in the argument, for non-smooth domains.
Let $U_j:=U(x,\alpha_0,k_j)$, $L:=\triangledown^2$, and let the overline
 denote complex conjugate. 
Then:
$$
I:=\int_{D_3}(U_1 L\overline {U_2} -\overline {U_2} LU_1)dx=
(k_1^2-k_2^2)\int_{D_3}
U_1\overline {U_2}dx .                 \tag 9
$$
 We wish to prove that the right-hand side vanishes. This follows if $I=0$.
The integral $I$ can be transformed formally by Green's formula and, using
 the boundary condition,
 one concludes that $I=0$. The problem is to justify the applicability of
 Green's formula in the
 domain $D_3$ with non-smooth boundary.
The remaining part of the proof contains such a justification.

Our starting point is the known (see [1],[2], [9]) result:

 Green's formula holds for the domains with finite perimeter and functions 
whose first derivatives
are in the space $BV$, provided that their rough traces are 
summable on the reduced boundary of the domain (in our case $D_3$ is 
the domain) with respect to
$(n-1)$-dimensional Hausdorff measure.

 Let $\Omega \in \Bbb R^n$ be a domain. Recall that the space $BV(\Omega)$
 consists of functions
 whose first derivatives are signed measures locally in
$\Omega$ ([2], [9]).
A set $D_j$ has finite perimeter if $\chi_j$, the characteristic function of
 this set, belongs to
 $BV(\Bbb R^n)$. The reduced boundary, denoted by $\Gamma^\ast$,
is the set of points at which the exterior normal in the sense of Federer
exists (see [1], [2], or [9] for the definition of this normal and [2] for
  that
of the  rough trace).
It is proved in [9], that for the sets with finite perimeter the 
reduced boundary has full $(n-1)$-dimensional Hausdorff measure, so that
the normal in the sense of Federer is defined almost everywhere on $\Gamma$
 with respect to  $(n-1)$-dimensional Hausdorff measure (we will write
 $s$-almost everywhere for brevity).
 What we need is to check that: 

1) the set $D_3$ has finite perimeter,

2) the function $\triangledown \cdot \psi$ is a measure in $D_3$, where 
 $ \psi:= U_1 \triangledown \overline {U_2}- \overline U_2 
\triangledown {U_1}$,

and

3) $\psi$ has a summable rough trace on $\Gamma^\ast_3$, the reduced boundary 
of $D_3$.

Note that the integrand in the first integral in formula (9) is of the form 
 $\triangledown \cdot \psi$, and  $\triangledown \cdot \psi=
 (k_1^2-k_2^2)U_1\overline U_2$.

First, let us prove that $D_3$ has finite perimeter. Note , that  $D_3$ 
is not necessarily a Lipschitz domain, although $D_1$ and $D_2$ are.
Let us denote by $P(D)$ the perimeter of $D$ and by $||\triangledown \chi||$
the norm of $\chi$ in the space $BV(\Bbb R^n)$, that is,  the  total
 variation of the vector 
measure $\triangledown \chi$. By definition,
 $P(D)=||\triangledown \chi||$.

Let $s(\Gamma)$ denote the $n-1$-dimensional Hausdorff measure of 
 $\Gamma$. It is known that $P(D)\leq s(\Gamma)$ and the 
strict inequality is possible for non-smooth  $\Gamma$. Also, it can 
happen that  $P(D)<\infty$,
but $s(\Gamma)=\infty$. If  $P(D)<\infty$, then $\Gamma^\ast$ is
$s$-measurable and $s(\Gamma^\ast)=P(D)$, see [9, p.193].

 The set $D_3$ has finite perimeter iff $||\triangledown \chi_3||<\infty$.
Clearly:
$$
\chi_3=\chi_{12}-\chi^{12},     \tag 10
$$
$$
\chi_{12}=\chi_1+\chi_2-\chi^{12},     \tag 11
$$
$$
\chi^{12}=\chi_1 \chi_2,     \tag 12
$$
where $\chi^{12}$, e.g., is the characteristic function of the  domain
$D^{12}$. By the assumption,  $||\triangledown \chi_j||<\infty,\, j=1,2$.
The space $BV$ is linear. Therefore, by formulas (10)-(12), it follows
that $P(D_3)<\infty$, if one checks that the function 
$\chi_1 \chi_2\in BV(\Bbb R^n)$. This, however, is a direct consequence of 
the known formula
 [9,p.189] for the derivative of the product of
bounded $BV$ functions: $\triangledown (\chi_1 \chi_2)=
\hat{\chi_2}\triangledown \chi_1 +\hat{\chi_1}\triangledown \chi_2$, where
 $\hat{\chi}$ denotes
 the averaged value of $\chi$ at
the point $x$ (see [9, p.189] for the derivation of this formula).
Note that the usual formula for the derivative of the product (the 
formula without the averaged values) is not valid for $BV$ functions,
in particular, it is wrong for the characteristic functions.

Let us now check that the function  $\triangledown \cdot \psi$ is a signed
 measure in $D_3$. Since 
$\triangledown \cdot  \psi= (k_1^2-k_2^2)U_1\overline U_2$ and the functions
 $U_1, U_2$ belong to
 $H^1(D_3)$, it follows that $U_1\overline U_2 \in L^1(D_3)$. Thus,
 $\triangledown \cdot \psi$ is a signed measure in $D_3$.

Finally, $\psi$ has a summable rough trace on $\Gamma^\ast_3$. In fact, more
 holds: the summable
 trace $\psi^+$ exists $s$-almost everywhere on $\Gamma^\ast_3$
and this implies existence of the summable rough trace. Recall that the
 trace  $\psi^+$ is defined
 at the point $x\in \Gamma$ as the following limit (if it exists):
$$
\psi^+(x)=\lim_{r\rightarrow 0} \frac 1{meas_n(D_r(x))}\int_{D_r(x)}\psi(y)dy,
$$
where $D_r(x):=\{y: y\in D_3,\, |x-y|<r\}$. Existence of the summable trace of
the function $\psi$ on $\Gamma_3^\ast$ follows from [10, lemma 5.7].
One can see that the trace exists in yet stronger sense: $U_j(x)$ and 
$\triangledown U_j(x)$ have non-tangential limits as 
$x\rightarrow t\in \Gamma_3^\ast$, these
 limits are in $L^2(\Gamma_3^\ast, ds)$ and therefore
their product is in $L^1(\Gamma_3^\ast, ds)$, that is , the trace of
 $\psi$ is summable.
This completes the proof of the uniqueness theorem for IOSP1. Let us
 formulate the result:
\proclaim{Theorem 1} Assume that the obstacles $D_j,\, j=1,2,$ have 
the following properties:

1) they are Lipschitz domains,

2)   $A_1(\alpha^\prime, \alpha_0,k)=A_2(\alpha^\prime, \alpha_0,k)\,\,
\forall \alpha^\prime\in S^{n-1},
\forall k\in [a,b], \, 0\leq a<b $.

Then $D_1=D_2$ and, in the case of Robin's boundary condition,
 $\sigma_1=\sigma_2$.
\endproclaim
\demo{Proof} Only the last statement is not yet proved. However, since we
 have already
 established that $D_1=D_2:=D$ and $u_1=u_2$ in $D^\prime$, it follows that
$$
\sigma_1=-\frac {u_{1N}}{u_1}=-\frac {u_{2N}}{u_2}=
\sigma_2 \,\,\text { on }\Gamma.\qed
$$
\enddemo

Another proof can be given. It is based on formula (13) and on the method,
 developed in section
 2.2 below. 

 If $\Gamma_j$ are Lipschitz boundaries, then the existence and uniqueness
 of the scattering
 solutions can be established as in [3] with the help of the potential
 theory for domains with
  Lipschitz boundaries [10]. The details of this theory will be published

 elsewhere.

In the next subsection we consider IOSP2 and use the method developed
in [5]-[7] for the uniqueness proof.
\subsubhead 2.2.Uniqueness for IOSP2\endsubsubhead 

The starting point is the identity first established in [5]:
$$
4\pi(A_1-A_2)=\int_{\Gamma_{12}}[u_1u_{2N}-u_{1N}u_2]ds,  \tag 13
$$
where $u_1:=u_1(x,\alpha,k),\, u_2:=u_2(x,-\alpha^\prime,k)$, $u_N$
denotes the normal derivative, as before,  $u_j$ and
$A_j:=A_j(\alpha^\prime,\alpha,k)$ are, respectively, the scattering
 solution and scattering
 amplitude, corresponding to the obstacle $D_j, \,j=1,2.$ 
Applications of this useful formula are given in [5]-[8].

If $A_1=A_2$ for the fixed energy  data in IOSP2, then (13) yields:
$$
0=\int_{\Gamma_{12}}[u_1(s,\alpha)u_{2N}(s,-\alpha^\prime)-
u_{1N}(s,\alpha)u_2(s,-\alpha^\prime)]ds, \quad 
\forall \alpha,\alpha^\prime \in S^{n-1}, \tag 14
$$
where we have dropped the dependence on the fixed energy $k_0$.

Let $G_j:=G_j(x,y,k)$ denote Green's function for the problem (1)-(3),
or the exterior problem with the Robin boundary condition. It is proved in
[3,p.46], that
$$
G_j=\gamma(r)[u_j(x,\alpha,k) +O(\frac 1r)],\quad r\to \infty,\,
 y/r:=-\alpha, \,r:=|y|, \tag 15
$$
where $\gamma(r)$ is a known function (e.g., $\gamma=
\frac {exp(ikr)}{4\pi r}$ 
if $n=3$), and  the coefficient $u_j$ in (15) is the scattering solution.
\proclaim{Lemma 1} Equation (14) implies:
$$
0=\int_{\Gamma_{12}}[G_1(s,x)G_{2N}(s,y)-G_{1N}(s,x)G_{2}(s,y)]ds,\quad \forall 
x,y\in D_{12}^\prime. \tag 16
$$
\endproclaim
\demo{Proof} We give a proof for $n=3$. For other $n$ the proof is similar.
First, let us derive the equation:
$$
W(y):=
\int_{\Gamma_{12}}[u_1(s,\alpha)G_{2N}(s,y)-u_{1N}(s,\alpha)G_{2}(s,y)]ds=0,
 \quad \forall 
y\in D_{12}^\prime, \forall \alpha\in S^{n-1}.\tag 17
$$
Indeed, $W(y)$ solves equation (1) in  $D_{12}^\prime$ and $W=o(1/r)$, as
 follows from (14) and (15).
Thus, $W=0$ in  $D_{12}^\prime$, see [3,p.25].

 Let us prove (16) now.
Fix any $y\in D_{12}^\prime$ and let $w$ denote the integral in (16).
Then $w$ solves (1) in  $D_{12}^\prime$ and $w=o(1/r)$, as follows from
(15) and (17). Thus, (16) follows and Lemma 1 is proved. $\qed$ 
\enddemo

We want to derive a contradiction from (16). This contradiction will prove 
that
$D_1=D_2$. According to the argument given in the section 2.1, the set
 $D_{12}$ has finite
 perimeter, Green's formula is applicable to (16) in the domain
 $D_{12}^\prime$, and we get
 the following equation:
$$
0=G_1(y,x)-G_2(x,y)  \quad \forall x,y\in D_{12}^\prime,   \tag 18
$$
where the radiation condition for $G_1$ and $G_2$ was used: it allowed us 
to neglect the integral
 over the large sphere, which appeared in Green's formula.

We now want to derive a contradiction from (18). Note that $G_j(x,y)=
G_j(y,x)$ and consider, for
 instance, the Neumann condition (2). The Robin condition is treated
 similarly. Differentiate
 (18) with respect to $y$ along the normal
$N_t, t\in \Gamma_2^\prime$, and let $y\rightarrow t$. This yields:
$$
0=G_{1N_t}(t,x)  \quad \forall x\in D_{12}^\prime,  t\in \Gamma_2^\prime.
 \tag 19
$$
The point $t$ belongs to $D_1^\prime$. Therefore
$$
|G_{1N_t}(t,x)|\rightarrow \infty \text { as } x\rightarrow t.  \tag 20
$$
Equation (20) contradicts (19). This contradiction proves that $D_1=D_2$.
We have proved the following result:
\proclaim{Theorem 2} Let the assumption 1)  of Theorem 1 hold and 
assume that

$2^\prime$)  $A_1(\alpha^\prime, \alpha,k_0)= A_2(\alpha^\prime,
 \alpha,k_0),\,\forall \alpha,\alpha^\prime\in S^{n-1}.$

Then $D_1=D_2$ and, in the case of Robin boundary condition,
 $\sigma_1=\sigma_2$.
\endproclaim

This completes the discussion of the uniqueness theorem for IOSP2 for the 
case of Neumann and Robin boundary conditions.

\vfill
 
\noindent e-mail: ramm\@math.ksu.edu

\enddocument